\documentclass[conference,10pt]{IEEEtran}
\IEEEoverridecommandlockouts
\usepackage{cite}
\usepackage{amsmath,amssymb,amsfonts}
\usepackage{soul}
\usepackage{algorithm}
\usepackage{algpseudocode}
\usepackage{graphicx}
\usepackage{textcomp}
\usepackage[font=small]{caption}
\usepackage{subcaption}
\usepackage{comment}
\usepackage{bm}
\usepackage{mathtools}
\def\BibTeX{{\rm B\kern-.05em{\sc i\kern-.025em b}\kern-.08em
    T\kern-.1667em\lower.7ex\hbox{E}\kern-.125emX}}

\newcommand{\calCN}{{\cal CN}}

\newcommand{\calK}{{\cal K}}
\newcommand{\calH}{{\cal H}}

\newcommand{\calM}{{\cal M}}

\newcommand{\calP}{{\cal P}}
\newcommand{\calS}{{\cal S}}
\newcommand{\calU}{{\cal U}}

\newcommand{\ds}{\displaystyle}

\newcommand{\ba}{\mathbf{a}}
\newcommand{\bb}{\mathbf{b}}
\newcommand{\bd}{\mathbf{d}}

\newcommand{\bh}{\mathbf{h}}
\newcommand{\bp}{\mathbf{p}}

\newcommand{\bs}{\mathbf{s}}
\newcommand{\bw}{\mathbf{w}}

\newcommand{\by}{\mathbf{y}}
\newcommand{\bz}{\mathbf{z}}

\newcommand{\bzero}{\mathbf{0}}

\newcommand{\bA}{\mathbf{A}}

\newcommand{\bC}{\mathbf{C}}
\newcommand{\bD}{\mathbf{D}}

\newcommand{\bF}{\mathbf{F}}
\newcommand{\bG}{\mathbf{G}}
\newcommand{\bH}{\mathbf{H}}
\newcommand{\bI}{\mathbf{I}}
\newcommand{\bP}{\mathbf{P}}

\newcommand{\bR}{\mathbf{R}}

\newcommand{\bU}{\mathbf{U}}
\newcommand{\bV}{\mathbf{V}}
\newcommand{\bW}{\mathbf{W}}

\newcommand{\bbE}{\mathbb{E}}
\newcommand{\bbC}{\mathbb{C}}

\newcommand{\test}{{\underset{H_0}{\overset{H_1}\gtrless}}}
\newcommand{\norm}[1]{\left\lVert#1\right\rVert}

\begin{document}

\onecolumn
\thispagestyle{empty}
\Huge IEEE Copyright Notice \\

\large
© 2025 IEEE. Personal use of this material is permitted. Permission from IEEE must be obtained for all other uses, in any current or future media, including reprinting/republishing this material for advertising or promotional purposes, creating new collective works, for resale or redistribution to servers or lists, or reuse of any copyrighted component of this work in other works.

\vfill
This work has been accepted for publication in \textit{2025 IEEE 26th International Workshop on Signal Processing and Artificial Intelligence for Wireless Communications (SPAWC)}. The final published version is available via IEEE Xplore, DOI: 10.1109/SPAWC66079.2025.11143520.

\clearpage
\twocolumn
\normalsize

\title{Power Control Design for ISAC Optimization in User-Target-Centric Cell-Free mMIMO Networks
    \thanks{This paper is an extension of the journal \cite{liesegang2025scalable} by the same authors. This work was carried out within the framework of the Horizon Europe Programme Marie Skłodowska-Curie Actions (MSCA) Postdoctoral Fellowships (PF) - European Fellowships (EF) DIRACFEC (Grant No. 101108043). The European Union (EU) also supported the work of Stefano Buzzi under the Italian National Recovery and Resilience Plan (NRRP) of NextGenerationEU, partnership on “Telecommunications of the Future” (PE00000001 - program “RESTART”, Structural Project 6GWINET, Cascade Call SPARKS). Views and opinions expressed are those of the author(s) only and do not necessarily reflect those of the EU. The EU cannot be held responsible for them.}
}

\author{\IEEEauthorblockN{Sergi~Liesegang$^{1,2}$ and Stefano~Buzzi$^{1,2,3}$}
\IEEEauthorblockA{$^1$\textit{DIEI, Università degli Studi di Cassino e del Lazio Meridionale, 03043 Cassino (FR) -- Italia} \\
$^2$\textit{Consorzio Nazionale Interuniversitario per le Telecomunicazioni, 43124 Parma (PR) -- Italia} \\ 
$^3$\textit{DEIB, Politecnico di Milano, 20122 Milano (MI) -- Italia}}
E-mails: \{sergi.liesegang, buzzi\}@unicas.it}

\maketitle

\begin{abstract}
This paper addresses the power control design for a cell-free massive MIMO (CF-mMIMO) system that performs integrated sensing and communications (ISAC). Specifically, the case where many access points are deployed to simultaneously communicate with mobile users and monitor the surrounding environment at the same time-frequency slot is considered. On top of the user-centric architecture used for the data services, a target-centric approach is introduced for the detection tasks. As a valuable performance metric, we derive the receive sensing signal-to-noise (SNR) ratio under generalized likelihood ratio test processing. Based on that, we formulate a quality-of-service (QoS) scheme that maximizes the two figures of merit: achievable data rate and effective sensing SNR. Simulations demonstrate that our proposal surpasses orthogonal resource algorithms, underscoring the potential of ISAC-enabled CF-mMIMO networks.
\end{abstract}


\section{Introduction} \label{sec:1}
6G networks will integrate sensing into their architecture, a concept known as integrated sensing and communications (ISAC), crucial for precise data transmission and situational awareness \cite{liu2022integrated}. ISAC uses wireless signals for both communication and sensing, enabling applications like autonomous vehicles, smart cities, and industrial automation. It transforms network capabilities by using the same transceiver and frequency for both tasks. Telecom providers can thus expand services beyond traditional uses to include advanced options like environmental surveillance and object tracking, creating new applications and revenue streams.

To integrate these innovative capabilities, the widespread deployment of ISAC systems necessitates sophisticated wireless network designs. A promising approach is the cell-free massive multiple-input multiple-output (CF-mMIMO) system \cite{demir2021foundations}, which transforms conventional cell-based architectures by eliminating fixed cell boundaries in favor of a distributed arrangement of access points (APs). In CF-mMIMO systems, numerous spatially scattered APs collaborate to deliver services to user equipments (UEs) utilizing shared spectrum resources. Typically, these APs have moderately-sized antenna arrays and are connected to one or more central processing units (CPUs)\footnote{The connectivity between the APs and the CPU can be realized through many technologies, including free space optics \cite{arapoglou2022variable}.}.

In the user-centric approach \cite{buzzi2019user}, a few APs serve each UE, enhancing energy efficiency, macro-diversity, and reducing latency and complexity. Combined with ISAC for surveillance, it enables distributed antenna systems for sensing. Target-centric architectures, using a subset of APs for each target, outperform unscalable systems \cite{liesegang2025scalable}. Also, CF-mMIMO allocates specific APs for echo reception, removing the need for full-duplex base stations, solving a key issue in ISAC with multi-cell massive MIMO \cite{liao2024powerallocation}.

Despite rapid advances, challenges persist in ISAC-enabled CF-mMIMO \cite{lu2024isac,meng2024cooperative}. In \cite{behdad2024isac}, target detection under channel estimation errors is studied. Demirhan et al. proposed a joint beamforming scheme to optimize the minimum spectral efficiency across users and sensing SNR \cite{demirhan2024cellfree}. The paper \cite{mao2024csregion} derived the communication-sensing (C-S) region. Most works assume a single target. This study investigates user-target-centric CF-mMIMO for multi-target ISAC and develops power control rules to enhance QoS, expanding on policies from \cite{liesegang2025scalable}, which focused on scalable but suboptimal algorithms. The paper specifically shows the superiority of ISAC approaches with respect to the case in which sensing and communication tasks adopt separate (orthogonal) resources. 

The remainder of this paper is organized as follows. Section~\ref{sec:2} describes the system model. Sections~\ref{sec:4} and \ref{sec:5} present the performance metrics, namely data rate and sensing SNR. Section~\ref{sec:4} derives the power control. Section~\ref{sec:6} provides the numerical experiments. Section~\ref{sec:7} concludes the work.

\section{System Model} \label{sec:2}
In this work, we consider a distributed ISAC-enabled CF-mMIMO deployment where $M$ APs transmit (i) data signals toward $K$ UEs and (ii) $S$ sensing beams for detecting potential targets. Unless otherwise stated, both tasks are performed simultaneously in the same (time-frequency) physical resource. APs are equipped with $N$ antennas, while UEs have single-antenna transceivers. The APs are connected via a high-capacity fronthaul link to a CPU, where the centralized system operations occur. For simplicity, it is assumed that all APs are connected to the same CPU; however, the case of multiple CPUs can be addressed as in \cite{bjornson2020scalable}.

We embrace the multi-static sensing approach \cite{behdad2024isac} and assign specific “receive” APs exclusively for gathering echoes from potential targets (which avoids the need for full-duplex APs). “Transmit” APs will then be responsible for communication tasks like uplink (UL) training and downlink (DL) data, plus sensing beams transmission. Accordingly, we denote the set of transmit and receive APs as $\cal{M}^{\rm tx}$ and $\calM^{\rm rx}$, respectively, such that $\vert \calM^{\rm tx} \cup \calM^{\rm rx} \vert = M$.

To ensure scalability in communication, we focus on \textit{user-centric} CF-mMIMO systems \cite{buzzi2019user}, where a subset of APs serves each UE. More precisely, let $\calM_k \subset \calM^{\rm tx}$ represent the set of APs serving UE $k$ and $\calK_m$ the set of UEs served by AP $m$. The UE-AP association rule is left unspecified until the simulations, to keep the analysis general.

\begin{figure}[t]
    \centerline{\includegraphics[scale = 0.36]{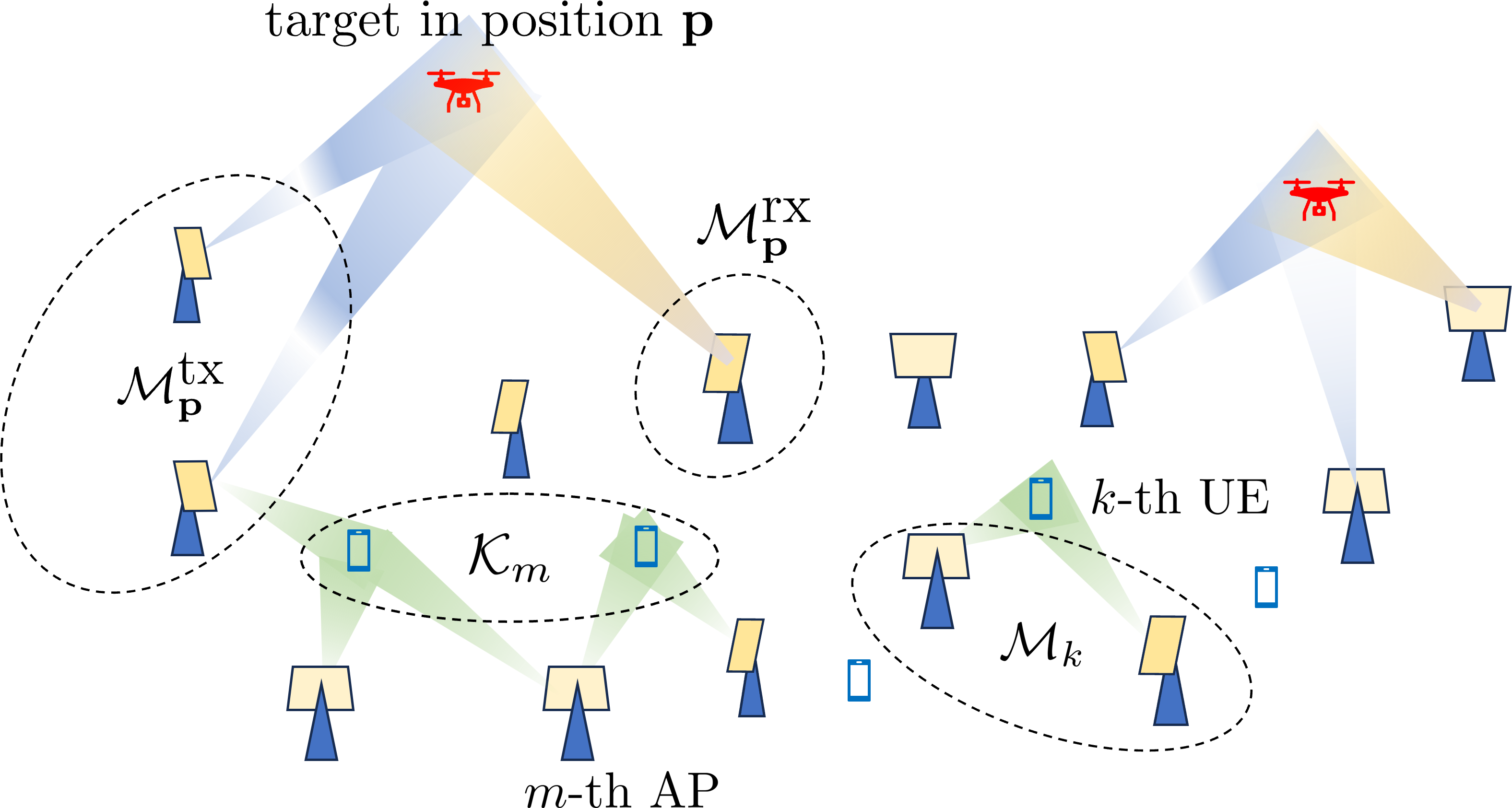}}
    \caption{Pictorial representation of the ISAC scenario under evaluation. Transmit/receive APs serve UEs and detect targets simultaneously.}
    \label{fig:1}
\end{figure}

Similarly, we support the \textit{target-centric} approach outlined in \cite{liesegang2025scalable}, wherein only selected APs participate in the detection of each target. Here, $\bp$ represents the spatial coordinates of the radar cell under observation, and $\calM_{\bp}$ defines the subset of APs tasked with detection. Specifically, $\calM_{\bp}^{\rm tx} \triangleq \calM_{\bp}\cap \calM^{\rm tx}$ and $\calM_{\bp}^{\rm rx} \triangleq \calM_{\bp}\cap \calM^{\rm rx}$ represent the subsets of transmit and receive APs, respectively. A visual example is shown in Fig.~\ref{fig:1}.

The surveillance area is equally split into $S$ non-overlapping \textit{sensing regions} scanned by each sensing beam, allowing $S$ cells to be examined concurrently, and preventing two neighboring cells from being inspected at the same time. We then denote by $\calS_m \subset \calS$ the subset of regions inspected by transmit AP $m$, with $\calS$ the total set such that $\vert \calS \vert = S$. Contrarily, receive APs will only be detecting a single target within their sensing region.
	
With these considerations, we focus on the detection at positions $\{\bp_i \, : \, i \in \calS\}$. Assuming block fading and OFDM modulation, each coherence block consists of $\tau_c$ time-frequency samples, in which channels and reflections are constant and flat. From now on, we use the index $t \in \{1,\ldots,\tau_c\}$ to denote the symbols available for communication and sensing \cite{liesegang2025scalable}.

\subsection{Propagation Channels}
The channel from UE $k$ to AP $m$ is $\bh_{k,m} \sim \calCN(\bzero,\bC_{k,m})$, with $\bC_{k,m} \in \bbC^{N \times N}$ the correlation matrix of the Rayleigh-distributed non-line-of-sight (NLoS) components.

The link between AP $m' \in \calM^{\rm tx}$ and AP $m \in \calM^{\rm rx}$ is $\bG_{m,m'} = \sqrt{1/(1 + c_{m,m'})}(\bar{\bG}_{m,m'} + \sqrt{c_{m,m'}} \bV_{m,m'})$, where $c_{m,m'}$ refers to the Rician factor \cite{3GPP36814}, $\bar{\bG}_{m,m'} \in \bbC^{N \times N}$ contains the correlated NLoS components, and $\bV_{m,m'} = e^{j\psi_{m,m'}} \ba_m(\varphi_{m,m'},\theta_{m,m'}) \ba_{m'}^{\rm H}(\varphi_{m',m},\theta_{m',m})$ is the array response at the LoS direction, with $\psi_{m,m'}\sim \calU[0,2\pi]$ is the phase offset and $\ba_m(\varphi_{m,m'},\theta_{m,m'})$ the steering vector for the azimuth $\varphi_{m,m'}$ ($\varphi_{m',m}$) and elevation $\theta_{m,m'}$ ($\theta_{m',m}$) angles of arrival (departure) from AP $m$ to AP $m'$.

Under LoS propagation between the target at position $\bp_i$ and the APs, let $\bA_{i,m,m'} = \ba_m(\phi_{m,i},\vartheta_{m,i}) \ba_{m'}^{\rm H}(\phi_{m',i},\vartheta_{m',i})$ so that the composite channel is $\bH_{i,m,m'} = \tilde{\alpha}_{i,m,m'} \bA_{i,m,m'}$, where $\tilde{\alpha}_{i,m,m'}=\alpha_{i,m,m'} \sqrt{\beta_{i,m,m'}}$, with $\alpha_{i,m,m'}$ the target reflectivity, or radar cross-section (RCS), and $\beta_{i,m,m'}$ the product of the path loss from transmit AP $m'$ to the target and that from the target to receive AP $m$. As mentioned, we follow the Swerling-I model for the RCS so that $\alpha_{i,m,m'}$ remains constant over consecutive symbols of the coherence block \cite{behdad2024isac}. Finally, $\phi_{m,i}$ $(\vartheta_{m,i})$ is the azimuth (elevation) angle of the position $\bp_i$ with respect to (w.r.t.) the antenna array of AP $m$.

\subsection{Signal Transmission}
Each transmit AP $m \in \calM^{\rm tx}$ sends data to the UEs in the set $\calK_m$ plus additional beams to detect the presence of potential targets within the surveillance area. The baseband equivalent of the $t$-th transmit signal with power $P_m$ is
\begin{equation}
    \begin{aligned}
        \bs_m[t] &= \sum\nolimits_{k \in \calK_m}\sqrt{\eta_{k,m}} \bw_{k,m} x_k [t] \\
        &\quad + \sum\nolimits_{i \in \calS_m} \sqrt{\mu_{i,m}} \bw_{0,m}(\bp_i) x_{i,m}[t],
    \end{aligned}
\end{equation}
where $\eta_{k,m}$ and $\mu_{i,m}$ are the power coefficients used for communication and sensing, respectively, such that $\sum_{k \in \calK_m} \eta_{k,m} + \sum_{i \in \calS_m} \mu_{i,m} \leq P_m$. $\bw_{k,m}$ and $x_k[t]$ are the unit precoder and the unit-energy data symbol intended for UE $k$, respectively. Similarly, $\bw_{0,m}(\bp_i)$ is the beamforming unit vector used to detect the potential target in position $\bp_i$, and $x_{i,m}[t]$ is a fake unit-norm data symbol associated with the $i$-th beam.

For the communication task, we will assume maximum ratio transmission, i.e., $\bw_{k,m} = \hat{\bh}_{k,m}/\sqrt{\bbE[\|\hat{\bh}_{k,m}\|^2]}$, with $\hat{\bh}_{k,m}$ the minimum mean-square error (MMSE) estimate of channel $\bh_{k,m}$ (whose expression is omitted due to space limitations but can be found in \cite[Subsection~III-B]{liesegang2025scalable}).

On the contrary, since we have perfect knowledge of the inspected position $\bp_i$, APs point the probing signal towards the angles $\phi_{m,i}$ and $\vartheta_{m,i}$. That is, the sensing beamformers are constructed as $\bw_{0,m}(\bp_i) = \ba_m(\phi_{m,i},\vartheta_{m,i})$.

\section{Achievable Data Rate} \label{sec:3}
The $t$-th received symbol of UE $k$ is given by
\begin{equation}
    y_k[t] = \sum\nolimits_{m \in \calM^{\rm tx}}\bh_{k,m}^{\rm H} \bs_m[t] + z_k[t],
\end{equation}
with $z_k[t] \sim \calCN(0, \sigma^2_z)$ the additive noise contribution. 

In cases of imperfect channel knowledge, the \textit{use-and-then-forget} bound provides a tractable rate expression, widely utilized in mMIMO literature. It is derived assuming that channel estimates are used solely for beamformer computation and discarded during signal detection.

Concisely, assuming the CPU has statistical information but no knowledge of the realizations, we have
\begin{align}
    y_k[t] &= \bar{\calH}_{k} x_k[t] + \left(\tilde{\calH}_{k} - \bar{\calH}_{k} \right) x_k[t] + z_k[t] \nonumber \\ \label{eq:3} &\quad + 
    \sum\nolimits_{j\neq k} \sum\nolimits_{m \in \calM_j} \sqrt{\eta_{j,m}} \bh_{k,m}^{\rm H}\bw_{j,m} x_j[t] \\
    &\quad +
    \sum\nolimits_{m \in \calM^{\rm tx}} \sum\nolimits_{i \in \calS_m} \sqrt{\mu_{i,m}} \bh_{k,m}^{\rm H} \bw_{0,m}(\bp_i) x_{i,m}[t], \nonumber
\end{align}
where the useful and uncertainty components are
\begin{equation*}
    \begin{aligned}
        \bar{\calH}_{k} &= \bbE\left[\sum\nolimits_{m \in \calM_k} \sqrt{\eta_{k,m}} \bh_{k,m}^{\rm H}\bw_{k,m}\right],\\
        \tilde{\calH}_{k} &= \sum\nolimits_{m \in \calM_k} \sqrt{\eta_{k,m}} \bh_{k,m}^{\rm H}\bw_{k,m}.
    \end{aligned}
\end{equation*}

In writing \eqref{eq:3}, we neglected the signals reflected by potential targets in the area since they are extremely weak and can be included in the multipath generating channels $\bh_{k,m}$.

Following the rationale in \cite{demir2021foundations}, we treat the resulting MMSE estimation errors as additional noise and consider the worst-case scenario (uncorrelated Gaussian model). This way, under the assumption of standard normal codebooks ($x_k[t],x_{i,m}[t] \sim \calCN(0,1)$, $\forall t$), a lower bound for the data rate reads as 
\begin{equation}
    R_k = \frac{\tau_c - \tau_p}{\tau_c} B \log_2\left(1 + \gamma_k\right),
\end{equation}
with $\tau_p$ the number of training symbols and $B$ the system's bandwidth. As shown in \cite[Section~IV]{liesegang2025scalable}, the communication signal-to-interference-plus-noise ratio (SINR) $\gamma_k$ is written as per \eqref{eq:5} at the top of the next page, where $\bm{\pi}_k \in \bbC^{\tau_p}$ is the pilot sequence and $\bm{\Phi}_{k,m} = \bm{\Lambda}_{k,m}\bC_{k,m}$ for convenience, with $\bm{\Lambda}_{k,m}$ the MMSE estimation matrix. Here, the sensing interference is modeled via the matrix $\bW_{i,m} = \bbE_{\calP_i}\left[\bw_{0,m}(\bp_i)\bw_{0,m}(\bp_i)^{\rm H}\right]$, where the expectation is w.r.t. the long-term distribution $\calP_i$ of the inspected locations $\bp_i$ along the data communication. This can be numerically calculated with Monte Carlo simulations.

\begin{figure*}[t]
    \begin{equation}
        \gamma_k = \frac{\ds \left\vert \sum\nolimits_{m \in \calM_k} \sqrt{\eta_{k,m} {\rm tr}\left(\bm{\Phi}_{k,m}\right)} \right\vert^2}{\ds \sum_{j = 1}^K\sum_{m \in \calM_j} \eta_{j,m} \frac{\ds {\rm tr}\left(\bC_{k,m} \bm{\Phi}_{j,m}\right)}{\ds {\rm tr}(\bm{\Phi}_{j,m})} + \sum_{j \neq k} \left\vert \bm{\pi}_j^{\rm H} \bm{\pi}_k \sum_{m \in \calM_j} \sqrt{\eta_{j,m}} \frac{\ds {\rm tr}\left(\bC_{k,m} \bm{\Lambda}_{j,m}\right)}{\ds \sqrt{{\rm tr}(\bm{\Phi}_{j,m})}} \right\vert^2 + \sum_{m \in \calM^{\rm tx}} \sum_{i \in \calS_m} \mu_{i,m} {\rm tr}\left( \bC_{k,m} \bW_{i,m} \right) + \sigma^2_z},
        \label{eq:5}
    \end{equation} 
    \hrule
\end{figure*}

\section{Sensing SNR} \label{sec:4}
Regarding detection, we will start by denoting by $\bar{\by}_m [t] \in \bbC^N$ the signal received by AP $m \in \calM_{\bp_i}^{\rm rx} = \calM_{\bp_i} \cap \calM^{\rm rx}$ to detect the possible presence of a target at position $\bp_i$ in the $i$-th detection region. We introduce a binary random variate, $b(\bp_i) \in \{0, 1\}$, which equals 1 if a target is present, and 0 otherwise. Accordingly, we end up with the observable
\begin{equation}
    \tilde{\by}_m[t] = \sum\nolimits_{i=1}^S b(\bp_i) \sum\nolimits_{m' \in \calM^{\rm tx}} \bH_{i,m,m'} \bs_{m'}[t] + \tilde{\bz}_m[t],
\end{equation}
with $\tilde{\bz}_m[t]\sim \calCN(\bzero, \sigma^2_z \bI_N)$ the thermal noise. In line with \cite{buzzi2024scalable}, we excluded the interference arising from the AP-AP links. The reason is that, since the CPU controls all APs in the system, it is reasonable to assume that the receiving AP knows the transmitted signal and that all channels $\bG_{m,m'}$ have been perfectly estimated. Thus, the corresponding components $\sum\nolimits_{m' \in \calM^{\rm tx}} \bG_{m,m'} \bs_{m'}[t]$ can be safely subtracted.

Based on the observations $\tilde{\by}_m [t]$, the network has to decide on the presence of targets in positions $\bp_1, \ldots, \bp_S$. Motivated by \cite{liesegang2025scalable}, we perform $S$ disjoint binary hypothesis tests to obtain a scalable solution. In short, for the $i$-th test of the target at position $\bp_i$, we consider the observables $\{\tilde{\by}_m [t] \, : \, m \in \calM_{\bp_i}^{\rm rx} \}$ and neglect the much weaker echo from potential targets at (far) positions $\bp_{i' \neq i}$. We then have
\begin{equation}
    \tilde{\by}_m[t] \approx b(\bp_i) \bD_{i,m}[t] \bm{\alpha}_{i,m} + \tilde{\bz}_m[t] \triangleq \tilde{\by}_{i,m}[t],
\end{equation}
where, to ease of notation, we introduced the target channel matrix $\bD_{i,m}[t] = [\bd_{i,m,1}[t], \ldots, \bd_{i,m,\vert\calM^{\rm tx}\vert}[t]]$, with columns $\bd_{i,m,m'}[t] = \sqrt{\beta_{i,m,m'}} \bA_{i,m,m'} \bs_{m'}[t]$, and the RCS vector $\bm{\alpha}_{i,m}=[\alpha_{i,m,1},\ldots, \alpha_{i,m,\vert\calM^{\rm tx}\vert}]^{\rm T}$.

After collecting $\tau_s \leq \tau_c$ sensing samples, we thus formulate the detection problem as the following binary hypothesis test
\begin{equation}
    \left\{\begin{array}{ll}
        H_1: & \ddot{\by}_{i,m} = \ds \ddot{\bD}_{i,m} \bm{\alpha}_{i,m} + \ddot{\bz}_m \\
        H_0: & \ddot{\by}_{i,m} = \ds \ddot{\bz}_m ,
    \end{array} \right.
    \label{eq:8}
\end{equation}
with $\ddot{\by}_{i,m} \in \bbC^{N \tau_s}$, $\ddot{\bD}_{i,m} \in \bbC^{N \tau_s \times \vert \calM^{\rm tx} \vert}$, and $\ddot{\bz}_m \in \bbC^{N \tau_s}$ the concatenation of $\tilde{\by}_{i,m}[t]$, $\tilde{\bD}_{i,m}[t]$, and $\tilde{\bz}_m[t]$, respectively. 

To solve \eqref{eq:8}, we model $\bm{\alpha}_{i,m}$ as an unknown but deterministic parameter\footnote{For the performance evaluation presented in Section~\ref{sec:6}, $\bm{\alpha}_{i,m}$ will be modeled as a correlated complex Gaussian vector (cf. \cite[Subsection~V-A]{liesegang2025scalable}).}. Unlike \cite{behdad2024isac}, this practice avoids having a detector structure tailored to a specific statistical model for the target reflectivity. As a result, we can resort to the generalized likelihood ratio test (GLRT), which amounts to substituting the maximum-likelihood estimate of $\bm{\alpha}_{i,m}$ in the likelihood ratio test and comparing it with a suitable threshold $\delta_i$ \cite{li2008mimo}: 
\begin{equation}
    \frac{\ds {\rm max}_{\bm{\alpha}_{i,m}} \, f\left(\ddot{\by}_{i,m} \vert H_1, \bm{\alpha}_{i,m}\right)}{\ds f\left(\ddot{\by}_{i,m} \vert H_0\right)} \test \delta_i.
\end{equation}

Notably, since when conditioning on $\bm{\alpha}_{i,m}$, the observable has a complex Gaussian distribution, it is easy to show that, when integrating at the CPU the contribution from all APs in the set $\calM_{\bp_i}^{\rm rx}$, the final GLRT can be written as
\begin{equation}
    \sum\nolimits_{m \in \calM_{\bp_i}^{\rm rx}} \norm{\bU_{i,m}^{\rm H}\bm{\Psi}_m^{-0.5}\ddot{\by}_{i,m}}^2 \test \ln \left(\delta_i\right),
    \label{eq:10}
\end{equation}
where $\bU_{i,m} \in \bbC^{N \tau_s \times r_{i,m}}$ contains the $r_{i,m}$ left-singular vectors of non-zero singular values from $\bm{\Psi}_m^{-0.5}\ddot{\bD}_{i,m}$ . Please refer to \cite[Section~V]{liesegang2025scalable} for more details on the complete proof.

Given the test \eqref{eq:10}, we can define the receive sensing SNR for the range cell at position $\bp_i$ in the $i$-th sensing area, say $\gamma_{\bp_i} $, as the ratio between the power of the useful signal in $\bU_{i,m}^{\rm H}\bm{\Psi}_m^{-0.5}\ddot{\by}_{i,m}$ and the power of its noise components, i.e.,
\begin{equation}
    \gamma_{\bp_i} = \frac{1}{r_i}\sum\nolimits_{m \in \calM_{\bp_i}^{\rm rx}} {\rm tr}\left(\bm{\Xi}_{i,m} \bbE\left[\bm{\alpha}_{i,m}\bm{\alpha}_{i,m}^{\rm H}\right] \bm{\Xi}_{i,m}^{\rm H}\right),
\end{equation}
with $r_i = \sum\nolimits_{m \in \calM_{\bp_i}^{\rm rx}}r_{i,m}$ and $\bm{\Xi}_{i,m} = \bU_{i,m}^{\rm H} \bm{\Psi}_m^{-0.5} \ddot{\bD}_{i,m}$ for brevity, and $\bR_{i,m}= \bbE[\bm{\alpha}_{i,m}\bm{\alpha}_{i,m}^{\rm H}]$ the RCS covariance matrix.

\section{Power Control Optimization} \label{sec:5}
The GLRT-based receive processing presented in Section~\ref{sec:4} is derived for a given power allocation, which means the performance metric $\gamma_{\bp_i}$ is obtained after $\eta_{k,m}$ and $\mu_{i,m}$ are tuned. To design these coefficients, one can obtain communication-like metrics for the sensing. In a nutshell, according to \cite{lu2024isac}, we can establish the (ergodic) mutual information over resource blocks as $\bar{R}_i = ( \tau_s/\tau_c) B \log_2(1 + \bar{\gamma}_i)$, with $\bar{\gamma}_i$ the “effective” SNR associated to hypothesis $H_1$ in \eqref{eq:8}:
\begin{equation}
    \bar{\gamma}_i = \frac{1}{N \tau_s \sigma_z^2}\ds \sum\nolimits_{m= 1}^{\vert \calM^{\rm tx} \vert} \bb_m^{\rm T} \bar{\bF}_{i,m} \bb_m,
\end{equation} 
where the matrix $\bar{\bF}_{i,m} \in \bbC^{KS \times KS}$ is given in \cite[Appendix]{liesegang2025scalable} and the vector $\bb_m$ contains the square-root power coefficients of transmit AP $m$, namely $\zeta_{k,m} = \sqrt{\eta_{k,m}},\nu_{i,m} = \sqrt{\mu_{i,m}}$.

\subsection{Coupled design}
With standard epigraph forms \cite{boyd2004convex}, we devise QoS-based power control rules for ISAC-enabled CF-mMIMO systems:
\begin{equation}
    \begin{aligned}
    \underset{\{\bb_m\}, \gamma}{\rm max} \, \gamma \quad {\rm s.t.} \quad &C1: \sum_{k \in \calK_m} \zeta_{k,m}^2 + \sum_{i \in \calS_m}\nu_{i,m}^2 \leq P_m, \, \forall m \\
    &C2: \bar{\gamma}_i \geq f(\gamma), \, \forall i, \quad C3: \gamma_k \geq g(\gamma), \, \forall k,
    \end{aligned}
    \label{eq:13} 
\end{equation}
where $C1$ constrains the power budget of each AP, while $C2$ and $C3$ guarantee a certain QoS for targets and UEs with the functions $f(\gamma)$ and $g(\gamma)$, respectively. Thanks to the previous formulation, the benefit is two-fold: one can either maximize the minimum (i) data rate subject to detection constraints, i.e., $g(\gamma) = \gamma$ and $f(\gamma) = \bar{\gamma}_0$, with $\bar{\gamma}_0$ a predefined threshold; or (ii) the “effective” sensing SNR subject to communication constraints, i.e., $f(\gamma) = \gamma$ and $g(\gamma) = \gamma_0$, with $\gamma_0$ fixed in advance. The former case has already been addressed in \cite[Subsection~VI-A]{liesegang2025scalable}, so we concentrate on the latter.

Like in most CF works, $C3$ is nonconvex since $\gamma_k$ is the ratio of convex functions. However, it can also be written as a second-order cone (SOC) constraint \cite{demirhan2024cellfree}:
\begin{equation}
    C3: \norm{\bm{\varrho}_k} \leq \sqrt{1 + \frac{1}{\gamma_0}} \sum_{m \in \calM_k} \zeta_{k,m} \sqrt{{\rm tr}\left(\bm{\Phi}_{k,m}\right)}, \quad \forall k,
\end{equation}
where the auxiliary variable $\bm{\varrho}_k$ follows from \cite[(50)]{liesegang2025scalable}.

Due to the SNR $\bar{\gamma}_i$ causing a nonconvex constraint $C2$, we use successive convex approximations to find a local optimum. For a given $\gamma$, the optimization is split into globally solvable subproblems. This involves approximating $\bar{\gamma}_i$ with a surrogate function to convexify $C2$ \cite{sun2017mm}. The optimal $\gamma$ is then found using a bisection search \cite{boyd2004convex}.

Among others, a common strategy is to linearize the function $\bar{\gamma}_i$ with the first-order Taylor expansions:
\begin{align}
    \bar{\gamma}_i\left(\bb\right) &\geq \bar{\gamma}_i\left(\bb^{(u - 1)}\right) + \nabla \bar{\gamma}_i\left(\bb^{(u - 1)}\right)^{\rm T}\left(\bb - \bb^{(u - 1)}\right) \nonumber \\
    &\triangleq \tilde{\gamma}_i\left(\bb,\bb^{(u - 1)}\right),
\end{align}
with $\bb$ the concatenation of all the “amplitude” coefficients, $u$ the iteration index, and the gradient
\begin{equation}
    \nabla \bar{\gamma}_i\left(\bb\right) = 2 \left(\sum\nolimits_{m= 1}^{\vert \calM^{\rm tx} \vert} \bP_m^{\rm T}\Re \lbrace \bar{\bF}_{i,m} \rbrace \bP_m\right) \bb,
\end{equation}
where $\bP_m \in \{0,1\}^{(K + S) \times M(K + S)}$ for notational consistency, i.e., $\bP_m \bb \equiv \bb_m$ (the coefficients from AP $m$).

Accordingly, at the $u$-th iteration (of each $\tilde{\gamma} \triangleq N \tau_s \sigma_z^2 \gamma$), we will need to solve the following feasibility problem:
\begin{equation}    
    \textrm{find} \quad \bb \quad \textrm{s.t.} \quad C1, C3, C2: \tilde{\gamma}_i\left(\bb,\bb^{(u - 1)}\right) \geq \tilde{\gamma}, \, \forall i.
\end{equation}

This way, we end up with a series of subproblems that are globally solved via standard numerical routines, such as CVX \cite{cvx2020}, and iterated until convergence to a stationary point.

\subsection{Independent design}
In the case of orthogonal resources (no ISAC), the previous problem leads to two separate optimizations:
\subsubsection{Communication} The optimal set $\{\zeta_{k,m}^{\star}\}$ follows from removing the sensing coefficients as well as constraint $C2$. As a result, \eqref{eq:13} is tackled without SCA methods.
\subsubsection{Sensing} The solution $\{\nu_{s,m}^{\star}\}$ is found by disregarding the communication coefficients and SOC $C3$.

\section{Numerical Simulations} \label{sec:6}
Throughout all experiments, we consider a deployment area of $0.5$ km\textsuperscript{2} equally divided into $S = 4$ sensing regions, where targets are uniformly deployed with heights ranging between $20$ m and $100$ m. UEs and APs are also randomly located but at fixed heights of $1.65$ m and $10$ m, respectively. We assume $K = 16$ UEs and $M = 16$ equipped with uniform linear arrays of $N = 4$ antennas. For simplicity, each UE is linked to the $4$ APs with the highest large-scale fading coefficients. Other association rules can be found in \cite{gennaro2024clustering}.

We employ the micro-urban scenario from \cite{3GPP36814}, with $P_m = 2$ W, $\sigma_z^2 = N_o B$, $N_o = -174$ dBm/Hz, and $B = 20$ MHz. The carrier frequency is $2$ GHz, and the coherence block has $\tau_c = 50$ symbols \cite{liesegang2025scalable}. The RCS is modeled as ${\alpha}_{i,m,m'} \sim \calCN(0,\sigma_{\alpha}^2)$, with variance $\sigma_{\alpha}^2 = 10$ dBsm (decibel relative to one square meter) and a Gaussian-shaped correlation based on the APs angle of view. The radar cells of each region are inspected by the closest receive AP and $4$ transmit APs.

\newpage

\begin{figure}[t]
    \centering    
    \includegraphics[scale = 0.95]{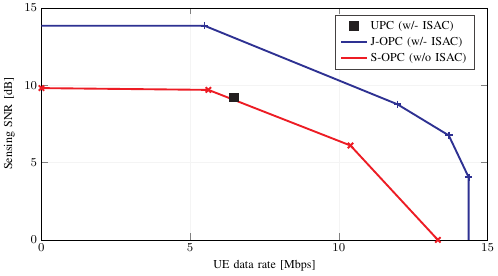}
    \vspace{-1mm}
    \caption{C-S region ($10\%$ quantile) w.r.t. power control rules. Points `$+$' are obtained by changing the QoS thresholds in \eqref{eq:13}, and points `$\times$' from changing the resources allocated to communication/sensing. \newline}
    \label{fig:2}
    \centering    
    \includegraphics[scale = 0.95]{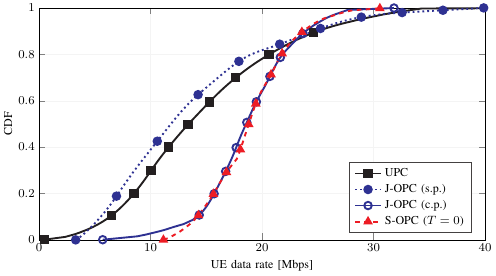}
    \vspace{-1mm}
    \caption{CDF of the UE data rate w.r.t. power control rules.}    
    \label{fig:3}
    \vspace{2mm}
    \centering    
    \includegraphics[scale = 0.95]{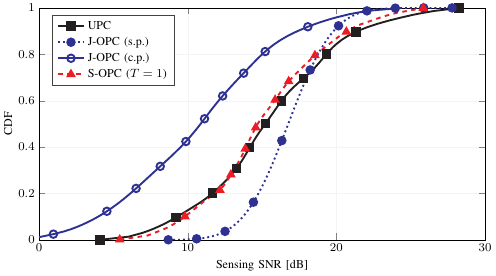}
    \vspace{-1mm}
    \caption{CDF of the sensing SNR w.r.t. power control rules.}
    \label{fig:4}
    \vspace{-3mm}
\end{figure}

In Fig.~\ref{fig:2}, we picture the C-S region w.r.t. the $10\%$-quantile for the coupled (or joint) optimal power control (J-OPC) and the independent (or separate) OPC (S-OPC). For comparison, we also include the uniform power control (UPC). The J-OPC points are obtained by repeatedly increasing the QoS threshold $\gamma_0$ (or $\bar{\gamma}_0$ \cite{liesegang2025scalable}) in problem \eqref{eq:13} until the optimization becomes unfeasible (cf. \cite{mao2024csregion}). The S-OPC boundary is computed by changing the ratio of physical resources dedicated to sensing, say $T = \tau_s/\tau_c \in [0,1]$, and UE communication, say $1 - T$. Remarkably, we see the joint setting outperforms the separate allocation. The data rates are practically equal only in the case of no sensing ($T = 0$). This proves the potential of ISAC.

The cumulative distribution functions (CDFs) of the data rate and sensing SNR are illustrated in Fig.~\ref{fig:3} and Fig.~\ref{fig:4}, respectively. Only two instances of J-OPC are shown to avoid overwhelming displays: the lower-right and the upper-left corners of the C-S region, denoted by \textit{communication-prioritized} (c.p.) and \textit{sensing-prioritized} (s.p.), respectively. In Fig.~\ref{fig:3}, the J-OPC (c.p.) and the S-OPC ($T = 0$) almost coincide. As expected, both surpass the UPC and the J-OPC (s.p.). However, note that the orthogonal-based approach leaves no room for sensing. In terms of QoS, ISAC is thus beneficial.

Fig.~\ref{fig:4} shows similar results, with J-OPC (s.p.) being the most effective. The best S-OPC $T = 1$ lags behind due to its inability to recycle communication signals, and the number of degrees of freedom (number of power coefficients) is also significantly reduced. This means ISAC is imperative.

\section{Conclusions} \label{sec:7}
The paper has discussed power control for CF-mMIMO systems with ISAC, using a user-target-centric architecture and GLRT processing. It has derived QoS policies to enhance UE data rate and sensing SNR. Experiments have shown the great advantages of ISAC w.r.t. orthogonal alternatives.

\bibliographystyle{IEEEtran}
\bibliography{references}

\end{document}